\documentclass[twoside]{article}
\usepackage{fleqn,espcrc2}

\newcommand{\AmS}{{\protect\the\textfont2
  A\kern-.1667em\lower.5ex\hbox{M}\kern-.125emS}}

\title{ Impact parameter dependence
 of parton densities at low $x$}

\author{S. M. Troshin and
 N. E. Tyurin \address{Institute for High Energy Physics,
Protvino, 142280 Russia}}
\begin{document}

\begin{abstract}
We consider impact parameter dependence of the
polarized and unpolarized parton densities.
 Unitarity does
not allow factorization  of these structure functions
over the Bjorken $x$ and the impact parameter $b$ variables.
On the basis of the
particular geometrical model approach, we conclude
that  spin of constituent
quark may have a significant orbital angular momentum
component which can manifest itself through the peripherality
of the spin dependent structure functions.\\
\vspace{1pc}
\end{abstract}


\maketitle
 The particular role of the low Bjorken $x$ region in DIS is
well known, it is this region where asymptotical properties
of the strong interactions can be studied. The characteristic
 property of this region is an essential contribution of
the nonperturbative dynamics \cite{bj,pasc} and one of the
possible ways to treat this dynamics  is the construction and
application of models.
Of course, the shortcomings of  any model approach
to the study of the nonperturbative dynamics are  evident.
However, one could hope to gain from a model approach an information
which cannot currently be obtained otherwise, e.g. by
 the lattice QCD methods.
The renewed topic  in the subject of structure functions at low $x$
  is the study of their
  geometrical features, i.e. the considerations of
the dependence  of the parton densities on the transverse
coordinates or the impact parameter.
 It has been
shown \cite{burk} that the $b$-dependent parton distributions are
related to the Fourier transform of the off-forward matrix
elements of parton correlation functions in the limiting case of
zero skewedness. Earlier connection of the form factors with the
$b$-dependent distributions was considered in \cite{chy} and
its generalization for the off-mass-shell case in \cite{petr}.

The study of the impact parameter  dependence
allows one to gain
information on the spatial distribution of the partons inside
the parent hadron and on the spin properties of the nonperturbative
intrinsic hadron structure. The geometrical properties of
structure functions  play also an important role
in the analysis of
the lepton--nuclei deep--inelastic scattering and in the
hard production in heavy--ion collisions.

To consider geometrical features of parton densities
 we suppose that the
DIS process is determined by the aligned-jet
mechanism \cite{bj}.  The aligned-jet mechanism
is  essentially nonperturbative. It
allows one to relate structure functions with the
discontinuities of the amplitudes
 of quark--hadron elastic scattering when an incoming quark is
 a  hadronlike object which is close to its mass shell.
These relations have the form \cite{jj}
 \begin{equation}
q(x, Q^2)  =
\frac{1}{2}\mbox{Im}[F_1(s,t,Q^2)+F_3(s,t,Q^2)]|_{t=0},
\end{equation}
\[
\Delta q(x,Q^2)
\]
\begin{equation}
\quad\quad\quad\;\;\; =\frac{1}{2}\mbox{Im}[F_3(s,t,Q^2)-F_1(s,t,Q^2)]|_{t=0},
\end{equation}
\begin{equation}\label{def1}
\delta q(x,Q^2) =  \frac{1}{2}\mbox{Im} F_2(s,t,Q^2)|_{t=0}.
\end{equation}

 The functions $F_i$ are  helicity amplitudes
for the elastic quark-hadron scattering in the standard notations
for the nucleon--nucleon scattering.
It should be noted that the structure functions obtained according
to the above formulas should be multiplied by the factor $\sim
1/Q^2$ --- probability that the virtual photon converts into
the asymmetric small-$p_\perp$
quark-antiquark pair \cite{bj}.
 The quark virtuality
 is connected to  the photon virtuality $Q^2$ and
this is reflected by the dependence of the functions $F_i$
on $Q^2$.
The amplitudes $F_i(s,t,Q^2)$ are the corresponding Fourier-Bessel
transforms of the functions $F_i(s,b,Q^2)$.
The relations (1-\ref{def1}) will be used as a starting
point for the definition of the impact parameter dependent
 structure functions, i.e. it is
natural to give the following operational  definition:
\begin{equation}
q(x,b,Q^2)  \equiv
\frac{1}{2}\mbox{Im}[F_1(x,b,Q^2)+F_3(x,b,Q^2)],
\end{equation}
\[
 \Delta q(x,b,Q^2)
\]
\begin{equation}
\quad\quad\quad\quad\;\;  \equiv
\frac{1}{2}\mbox{Im}[F_3(x,b,Q^2)-F_1(x,b,Q^2)],
 \quad
\end{equation}
\begin{equation}\label{def2}
 \delta q(x,b,Q^2) \equiv  \frac{1}{2}\mbox{Im}F_2(x,b,Q^2),
\end{equation}
and $q(x,Q^2)$, $\Delta q(x,Q^2)$ and $\delta q(x,Q^2)$ are the integrals
over $b$ of the corresponding $b$-dependent distributions, i.e.
\begin{eqnarray}
q(x,Q^2)&  = & \frac{Q^2}{\pi^2 x}\int_0^\infty db \,b\,q(x,b,Q^2),\\
\Delta q(x,Q^2)&  = & \frac{Q^2}{\pi^2 x}\int_0^\infty db\,b\,\Delta q(x,b,Q^2),\\
\label{int1}
\delta q(x,Q^2)&  = & \frac{Q^2}{\pi^2 x}\int_0^\infty db\,b\, \delta q(x,b,Q^2).
\end{eqnarray}

The functions $q(x,b,Q^2)$, $\Delta q(x,b,Q^2)$ and $\delta q(x,b,Q^2)$
  have  simple  interpretation, e.g.
the function $q(x,b,Q^2)$ is a number density of quarks $q$ with
fraction $x$ of the hadron longitudinal momentum at the transverse
distance $b$ from the hadron geometrical center. It should be
noted that unitarity plays crucial role in the probabilistic
interpretation of the function $q(x,b,Q^2)$. Indeed due to
unitarity
\[
0\leq q(x,b,Q^2)\leq 1.
\]
The integrated distribution $q(x,Q^2)$  is not limited by unity
and can have arbitrary non--negative value.

 Interpretation of
the spin distributions directly follows from their definitions:
they are the differences of the corresponding spin dependent quark
number densities.

Unitarity can be fulfilled through the $U$--matrix representation
for the helicity amplitudes of elastic quark--hadron scattering.
In the impact parameter representation the expressions for the
helicity amplitudes are the following
\begin{equation}\label{f1u}
F_{1,3}(x,b,Q^2)  =\frac { U_{1,3}(x,b,Q^2)}{[1-iU_{1,3}(x,b,Q^2)]},\quad
\end{equation}
\begin{equation}
F_2(x,b,Q^2)  = \frac {U_2(x,b,Q^2)}{[1-iU_1(x,b,Q^2)]^2}.\label{fu}
\end{equation}
Unitarity requires Im $U_{1,3}(x,b,Q^2)\geq 0$. It is
to be noted that the $U$--matrix
 form of the unitary
representation, contrary to the eikonal one, does not generate
 essential singularity in the complex $x$ plane
 at $x\to 0$.

We consider
 the structure functions
along the lines outlined in~\cite{usdif}.
A hadron consists of the constituent quarks aligned in the
longitudinal direction and embedded into the nonperturbative
vacuum (condensate). The constituent quark appears
as a quasiparticle, i.e. as current valence quark surrounded by
the cloud of quark-antiquark pairs of different flavors.
 We refer to effective QCD approach and  use the Nambu--Jona-Lasinio (NJL) model
 as a basis. The  Lagrangian  in addition to the
four--fermion interaction ${\cal L}_4$ of the original NJL model includes
the six--fermion $U(1)_A$--breaking term
${\cal{L}}_6\propto K(\bar u u)(\bar d d)(\bar s s)$.
Transition to partonic picture
 is described by the introduction of a momentum cutoff
 $\Lambda=\Lambda_\chi\simeq 1$ GeV, which corresponds to the scale
of chiral symmetry spontaneous breaking.
This  picture for a hadron structure implies  that  overlapping  and
 interaction of peripheral condensates in hadron  collision  occurs
 at the first stage. In the overlapping region the condensates
 interact and as a result virtual massive quark pairs appear.
  A part  of  hadron  energy  carried  by  the peripheral
 condensates goes to generation of massive quarks. In other words
 nonlinear field couplings  transform  kinetic  energy  into  internal
 energy  of dressed quarks. Of course,
 number of such  quarks fluctuates.  The average number of quarks in
 the considered case is
 proportional  to the convolution  of the condensate distributions
  $D^{Q,H}_c$
 of the colliding constituent quark and hadron:
 \begin{equation} N(s,b,Q^2) \simeq N(s,Q^2)\cdot D^Q_c
 \otimes D^H_c, \end{equation} where the function  $N(s)$  is
 determined  by  a transformation thermodynamics    of  kinetic
 energy  of interacting  condensates to the internal energy of
 massive quarks. To estimate the $N(s,Q^2)$ it is feasible to assume that
 it is proportional to the  maximal possible energy dependence
 \begin{equation} N(s,Q^2) \simeq \kappa(Q^2)  {(1-\langle x_Q\rangle )
 \sqrt{s}}/{\langle m_Q\rangle}, \label{kp} \end{equation} where
  $\langle x_Q \rangle
 $ is the average fraction of energy carried by the constituent quarks,
 $\langle m_Q\rangle $
 is the mass scale of constituent quarks.
In the model each of the constituent valence quarks located in the
 central part of the hadron
 is supposed to scatter in a quasi-inde\-pen\-dent way by the
 generated virtual  quark pairs at given impact parameter and by  the
  other valence quarks.
 The strong
interaction radius of the constituent quark $Q$ is determined
by its Compton wavelength and
the $b$--dependence of the function $\langle f_Q \rangle$ related to
  the quark form factor $F_Q(q^2)$ has a simple form $\langle
 f_Q\rangle\propto\exp(-m_Qb/\xi )$.
The helicity flip transition, i.e. $Q_+\rightarrow Q_-$
 occurs when the valence   quark
knocks out  a quark with the opposite  helicity
and the same  flavor.
The helicity functions
$U_i(x,b,Q^2)$
 at small values of $x$  have the following
  dependence  \cite{usdif}:
\begin{eqnarray}
U_{1,3}(x,b,Q^2) & = & c_{1,3}(x,Q^2)U_0(x,b,Q^2),\\
U_2(x,b,Q^2) & = & d(x,b,Q^2)U_0(x,b,Q^2),
\label{uis}
\end{eqnarray}
where
\[
c_{1,3}(x,Q^2)=1+{\beta _{1,3}(Q^2)m_Q \sqrt{x}}/{Q},
\]
\[
d(x,b,Q^2) =  \frac{g_f^2(Q^2)m_Q^2x}{Q^2}
\exp[-{ 2(\alpha -1)m_Qb}/{\xi}]
\]
and
\begin{eqnarray}
 U_0(x,b,Q^2) & = & i\tilde U_0(x,b,Q^2)\nonumber \\
&  = & i\left[\frac{a(Q^2)Q}{m_Q\sqrt{x}}\right]^{N}\exp[-Mb/\xi]
\label{u0}.
\end{eqnarray}
  $a$, $\alpha$, $\beta$, $g_f$ and $\xi$ are the model
parameters, some of them  depend on the virtuality $Q^2$.

Then using Eqs. (\ref{f1u},\ref{fu}) we obtain at small
$x$:
\begin{equation} \label{qxb}
q(x,b,Q^2)  = \frac{\tilde U_0(x,b,Q^2)}{1+\tilde
U_0(x,b,Q^2)},\quad
\end{equation}
\begin{equation}
\Delta q(x,b,Q^2) = \frac{c_-(x,Q^2)}{2}\frac{\tilde U_0(x,b,Q^2)}
{[1+\tilde U_0(x,b,Q^2)]^2},
\end{equation}
\begin{equation} \label{dqxb}
\delta q(x,b,Q^2) =  \frac{d(x,b,Q^2)}{2}
\frac{\tilde U_0(x,b,Q^2)}{[1+\tilde U_0(x,b,Q^2)]^2},
\end{equation}
where $c_-(x,Q^2) = c_3(x,Q^2)-c_1(x,Q^2)$.

From the above expressions it follows that $q(x,b,Q^2)$ has a central
$b$--dependence, while $\Delta q(x,b,Q^2)$ and $\delta q(x,b,Q^2)$ have
peripheral profiles.

From Eqs.~(\ref{qxb}-\ref{dqxb}) it follows that factorization
of $x$ and $b$ dependencies is not allowed by unitarity and this
provides certain constrains for the model parameterizations of structure
functions. Indeed, it is clear
 from Eqs.~(\ref{qxb}-\ref{dqxb}) that factorized form of the input
  ``amplitude'' $\tilde U_0(x,b,Q^2)$ cannot survive after unitarization
   due to the presence of the denominators.
It is to be noted here that from the relation
of impact parameter distributions with the off-forward parton
distributions  \cite{burk}
it follows that the same conclusion on the absence of
factorization is also valid for the off-forward parton distributions
with zero skewedness.

It is interesting to note that the spin structure functions
have a peripheral dependence on the impact parameter contrary
to  the central profile of the unpolarized structure function.
It could be related with the orbital angular momentum of quark pairs
inside the constituent quark.
The important point  is what the origin of this orbital angular momenta
 is.
It was proposed to use an analogy with
 an  anisotropic extension of the theory of superconductivity
 which seems to match well with the  picture for a constituent
  quark.  The studies of that theory show that the
   presence of anisotropy leads to axial symmetry of pairing
  correlations around the anisotropy direction $\hat{\vec{l}}$ and to
 the particle currents induced by the pairing correlations.  In
 other words it means that a particle of the condensed fluid
   is surrounded
 by a cloud of correlated particles  which rotate around
it with the
axis of rotation $\hat {\vec l}$.
 Calculation of the
orbital momentum  shows that it is proportional to the density of the
correlated particles.
Thus, it is clear that there is a direct analogy
between  this picture and the one describing the constituent quark. An
axis of anisotropy $\hat {\vec l}$ can be associated with the
polarization vector of current valence quark located at the origin of the
constituent quark.  The orbital angular momentum $\vec L$ lies
along $\hat {\vec l}$.

The spin of a constituent quark, e.g. $U$-quark, in the used model
 is given
by the  sum:
\begin{eqnarray}
J_U  = 1/2 & = & S_{u_v}+S_{\{\bar q q\}}+L_{\{\bar q q\}}\nonumber\\
& = & 1/2+S_{\{\bar q q\}}+L_{\{\bar q q\}}\label{bal}.
\end{eqnarray}
In principle, $S_{\{\bar q q\}}$ and $L_{\{\bar q q\}}$
 can include contribution of
gluon angular momentum. However,
since we consider effective Lagrangian approach where gluon
 degrees of freedom are overintegrated, we do not touch
the  problems of the
 separation and mixing of the quark  angular momentum and
 gluon effects in QCD.
 Indeed, in the extension of the  NJL--model
 the six-quark
 fermion operator simulates the effect of gluon operator
$\frac{\alpha_s}{2\pi} G^a_{\mu\nu}\tilde G^{\mu\nu}_a$,
where $G_{\mu\nu}$ is
the gluon field tensor in QCD.

The value of the orbital
 momentum contribution into the spin of constituent quark can be
 estimated according to  the relation between
 contributions of current quarks into a proton spin and corresponding
contributions of current quarks into a spin of the constituent quarks
and that of the constituent quarks into  proton spin:
\begin{equation} (\Delta\Sigma)_p = (\Delta U+\Delta
D) (\Delta\Sigma)_U,\label{qsp} \end{equation}
where $(\Delta\Sigma)_U=S_{u_v}+S_{\{\bar q q\}}$.
The value of $(\Delta\Sigma)_p$ was measured in the deep--inelastic
scattering.
Thus, on the grounds of the experimental data for polarized DIS
we arrive to conclusion that the significant part of the spin
of constituent quark  should be associated with
 the orbital angular momentum
of the current quarks inside the constituent one.

Then the peripherality of the spin structure functions
 can be correlated with the large contribution of the
 orbital angular momentum,
   i.e. with the  quarks
  coherent rotation.
Indeed, there is a compensation between the total spin
of the quark-antiquark cloud and its
  orbital angular momenta, i.e.
$L_{\{\bar q q\}}=-S_{\{\bar q q\}}$ and therefore
   the above correlation follows
 if such compensation has a local nature
 and valid for a fixed impact parameter.

The important role of orbital
 angular momentum was known \cite{sig}
 before  the European Muon Collaboration at CERN
 found that only small fraction of a nucleon spin is due to the quark
 contribution  and reappeared later
 as  one of the  transparent explanations of the polarized
  deep-inelastic scattering data \cite{ell}. Lattice QCD calculations
  in the quenched approximation also indicate significant quark orbital
  angular momentum contribution to  spin of a nucleon \cite{math}.
  The issue of the orbital angular momentum is important for the
  explanation of
polarization effects in the hyperon productions \cite{hyp}.

\bf {Acknowledgements}\rm: We are grateful to V.A.~Petrov for the
useful discussions. One of the authors (S.T.) would like to thank
Organizing Committee of the Workshop Diffraction 2000 in Cetraro,
Italy  for the invitation, financial support and exciting
scientific atmosphere in this beautiful place.   This talk is
based on the work supported in part by the Russian Foundation for
Basic Research under Grant No. 99-02-17995.

\end{document}